\documentclass[%
 reprint,
superscriptaddress,
 amsmath,amssymb,
 aps,
 prl,
 showkeys,
floatfix
]{revtex4-2}

\usepackage{graphicx}
\usepackage{dcolumn}
\usepackage{bm}
\usepackage{svg}
\usepackage{placeins}
\usepackage{float}
\usepackage{xcolor}
\usepackage{tikz}
\usepackage{caption}
\usepackage{upgreek}
\usepackage[T1]{fontenc}

\usepackage{romannum}
\AtBeginDocument{\pagenumbering{arabic}}
\usepackage[super]{nth}
\usepackage[newcommands]{ragged2e}
\usepackage{siunitx}
\DeclareUnicodeCharacter{2009}{\,} 
\DeclareUnicodeCharacter{3B1}{\ensuremath{\alpha}}
\DeclareSIUnit\angstrom{\text {Å}}

\usetikzlibrary{plotmarks}

\begin{document}

\preprint{APS/123-QED}

\title{Molecular Rheology of Nanoconfined Polymer Melts}% Force line breaks with \\
\author{Ahmet Burak Yıldırım}
\affiliation{Department of Mechanical Engineering, Bilkent University, 06800 Ankara, Turkey}

\author{Aykut Erbaş}
\affiliation{UNAM - National Nanotechnology Research Center and Institute of Materials Science \& Nanotechnology, Bilkent University, 06800 Ankara, Turkey}
\affiliation{Institute of Physics, University of Silesia, Katowice, Poland}

\author{Luca Biancofiore}
\affiliation{Department of Mechanical Engineering, Bilkent University, 06800 Ankara, Turkey}
\affiliation{UNAM - National Nanotechnology Research Center and Institute of Materials Science \& Nanotechnology, Bilkent University, 06800 Ankara, Turkey}

\date{\today}% It is always \today, today,

\begin{abstract}
We use non-equilibrium atomistic molecular dynamics simulations of unentangled melts of linear and star polymers ($\mathrm{C_{25}H_{52}}$) to study the steady-state viscoelastic response under confinement within nanoscale hematite $\left ( \mathrm{\alpha-Fe_2O_3} \right )$ channels. We report (i) the negative (positive) first (second) normal stress difference and (ii) the presence of viscoelastic tension at low shear rates. We link these effects to bond alignment such that chains near the surface can carry the elastic force exerted on the walls, which decays as the chains become more aligned in the flow direction as the shear rate increases.
\end{abstract}

%\keywords{Polymer melts, molecular dynamics, viscoelasticity, normal stress differences, relaxation time, confined geometry, bond orientation tensor} 
\maketitle
Nanoscale confinement of polymeric liquids is often encountered at ultra-thin fluid film lubrication and nano-manufacturing \cite{nezbeda_industrial_2021, bhushan_springer_2010}.
%wherein control over the exerted load on the surfaces and phase transitions of the liquid possess high importance \cite{ahmed_new_2021,gamaniel_effect_2021,paggi_modeling_2020,shaw_introduction_nodate,nafar_sefiddashti_flow-induced_2020}. 
%As the distance between surfaces confining the polymeric liquid, $h$, approaches the dimensions of the constituting molecules $R_{\mathrm{G}}$, the flow dynamics and corresponding viscoelastic response deviate from what continuum models predict~\cite{todd_nonequilibrium_2017,savio_multiscale_2015}. 
As the confinement of a solvent-free polymeric liquid (i.e., melt), $h$, approaches the dimensions of the constituting polymer chains (e.g., the radius of gyration $R_{\mathrm{G}}$), effects that are otherwise not visible in bulk can influence the rheological properties of the liquid~\cite{todd_nonequilibrium_2017, savio_multiscale_2015}. 
%This is partly because the increasing confinement can significantly alter the  dynamics of  chains, which determines the flow response of polymeric liquids along with the collective behaviour of the  chains composing the liquid~\cite{kavokine_fluids_2021, harasim_direct_2013}.
%As the confinement of a solvent-free polymeric liquid (i.e., melt) approaches the molecular dimension of the constituting polymer chains, effects that are otherwise not visible in bulk can influence the rheological properties of the liquid. 
%
For instance, as $h/R_{\mathrm{G}} \rightarrow 1$, polymer chains can adjust their configuration to the confinement by forming well-defined and discrete molecular layers, and these layers can alter the shear response of the melt in comparison to bulk~\cite{zheng_molecular_2013, drummond_dynamic_2000, mate_tribology_2008,yamada_general_nodate, smith_solidification_2019,israelachvili_intermolecular_nodate}. Further, confining surfaces can directly interact and induce non-negligible forces as the surfaces approach each other \cite{derjaguin_structural_1974, lifshitz_theory_1992}. 
%At strong confinement (e.g., $h/R_{\mathrm{G}} \approx 1$), the disordered molecules can also form ordered layers under shear and intensify the anisotropy of the stress response \cite{zheng_molecular_2013, drummond_dynamic_2000, mate_tribology_2008,yamada_general_nodate, smith_solidification_2019,israelachvili_intermolecular_nodate}. 
Additionally, the molecular variables, such as polymerization degree of chains, branching, and composition, can play roles in chain dynamics differently compared to bulk \cite{karatrantos_modeling_2019}. Thus, describing the rheological properties of polymer melts at nanoscale confinement requires alternative methods that can take care of phenomena arising from finite-system dimensions.

%One example is the appearance of well-defined layers of molecules depending on the film thickness and the effective diameter of the molecules, which is explained by the oscillatory and exponentially decaying solvation pressure as $h$ increases \cite{smith_solidification_2019, mate_molecular_1991, gao_structures_nodate,lim_solvation_2002}. The repulsion of surfaces is observed when $h$ is a multiple of $d$, while, vice versa, attraction is present. Also, within nanoconfined channels, a direct interaction between the opposing surfaces becomes non-negligible, which is governed by the van der Waals forces and characterized by the Hamaker constant of the two surfaces. This effect rapidly decreases as $h$ increases . . Emerging entanglements and reactive groups influence the rheological behaviour and often enhance the viscosity of the polymer solution \cite{rudisill_non-equilibrium_1993, anwar_nonlinear_2019, shaw_introduction_nodate}.

While continuum models rely on constitutive equations to establish a relationship between deformation and the stress tensor by introducing response functions (i.e., moduli)~\cite{spagnolie_introduction_2015}, they fall short of capturing the rheological response as one of the dimensions of the system becomes comparable to $R_G$ \cite{kirby_nodate, ahmed_new_2021, gamaniel_effect_2021, paggi_modeling_2020, shaw_introduction_nodate, nafar_sefiddashti_flow-induced_2020}. 
%At this level of confinement, the boundary layer that is formed by the chains adsorbed to the surfaces becomes comparable with $h$; hence the conformation of chains near the confining surfaces significantly influences the chain dynamics and the viscoelastic response \cite{ewen_slip_2018, napolitano_irreversible_2020, krim_friction_2012}. 
Furthermore, a microscopic definition of the stress tensor  can describe the viscoelastic response of melts if all the forces and coordinates of constituting atoms are available~\cite{todd_nonequilibrium_2017,doi1988theory}, for instance, from molecular simulations. Nevertheless, such an approach can be computationally expensive as the system dimensions become larger or when longer sampling times are needed. 
Alternatively, leveraging the contributions of continuum and molecular models, the complex rheological response of polymer melts across different length scales  can be captured~\cite{rudisill_non-equilibrium_1993, holland_enhancing_2015, morozov_introductory_2007, carreau_rheological_1972, laso_calculation_1993}.

In this work, we attempt to bridge molecular-scale phenomena and macroscopic behaviour of strongly-confined (i.e.,  nanoscale) polymeric melts under non-equilibrium conditions. By using extensive all-atom molecular dynamics (MD) simulations, we demonstrate that the microscopic stress tensor under steady-state shear deviates significantly from what the continuum model of a simple generalized Newtonian fluid (GNF) would predict. Our calculations show that these deviations lead to excess viscoelastic stress, which decreases as the confinement decreases ($h\gg R_{\mathrm{G}}$) and eventually vanishes for bulk systems for a wide range of molecular conditions (e.g., chain topology and rigidity) and confinement levels. 
%We  quantify the steady-state viscoelasticity without referring to macroscopic measures (e.g., storage and loss moduli), which are often used in literature \cite{oakley_molecular_1998, poh_universality_2020, roland_determining_2004, ewen_advances_2018}. 
We further relate the excess stress with the spatial orientation of the
carbon-carbon bonds of the chains  near the surface. Unexpectedly, chains adsorbed on the surface lead to an effective polymer brush-like layer, further decreasing the shear response~\cite{Erbas2015,Semenov:1995gg}.

%We  explain the deviation in terms of the microscopic observables (i.e., the bond-orientation tensor)  

%\textbf{Specifically, we use non-equilibrium MD simulations to shear polymer melts to address  (i) how the viscoelastic effects emerge under nanoscale confinement, (ii) how viscoelasticity differs from bulk behaviour, and (iii) how the chain rigidity of different polymer topologies (i.e., linear and star polymers) affects the steady-state viscoelastic response. }
%
%We reveal the dependence of relaxation time on the chain rigidity, such that, by varying the chain rigidity, we modify the contribution of the elastic forces without altering the molecular structure \cite{everaers_rheology_2004}.
%
%and we further discuss the molecular origins of viscoelasticity in polymer melts by exploiting the bond-orientation tensor that we define. 
%To this end, we use molecular dynamics (MD) simulations, which are recognized as a powerful tool for studying rheology at the nanoscale \cite{ewen_nonequilibrium_2017, rudisill_non-equilibrium_1993, jabbarzadeh_effect_2002, markesteijn_comparison_2012, kondratyuk_probing_2020}.

All MD simulations were performed using L\textsc{ammps} MD package \cite{thompson_lammps_2022} (see the Supplemental
Material for the simulation details). The interactions between the atoms are modelled by the standard 12/6 Lennard-Jones potential using D\textsc{reiding} force field \cite{mayo_dreiding_1990, das_solvation_1996}. Each simulation box contains at least $N=128$ charge-neutral linear (pentacosane) or star (branched 3-arm, 9-octylheptadecane) $\mathrm{C_{25} H_{52}}$ polymers equilibrated at $T = 372$~K at a pressure at $p = 1$ atm ~\cite{mccabe_examining_2001}. We control the chain rigidity through an angle  potential penalizing chain bending~\cite{everaers_rheology_2004, auhl_equilibration_2003} as $U_{\mathrm{angle}}\left ( \theta \right ) = k_{\theta} \left ( \theta - \theta_0 \right )^2$, where $\theta$ is the angle between triplets of atoms in a chain, $\theta_0 = 109.47^\circ$ is the equilibrium value of the angle \cite{callister_materials_nodate, das_solvation_1996}, and $k_{\theta} \equiv \epsilon = 50$ kcal/mol is the potential strength \cite{mayo_dreiding_1990}. By varying $k_{\theta}$, we obtain chains with varying relaxation dynamics~\cite{kalathi_rouse_2014, ripoll_star_2006, halverson_rheology_2012, auhl_equilibration_2003} and modify the contribution of the elastic forces without altering the molecular structure \cite{everaers_rheology_2004} for both polymer topologies.

\begin{figure}[b]
    \centering
    \includegraphics[width=0.475\textwidth]{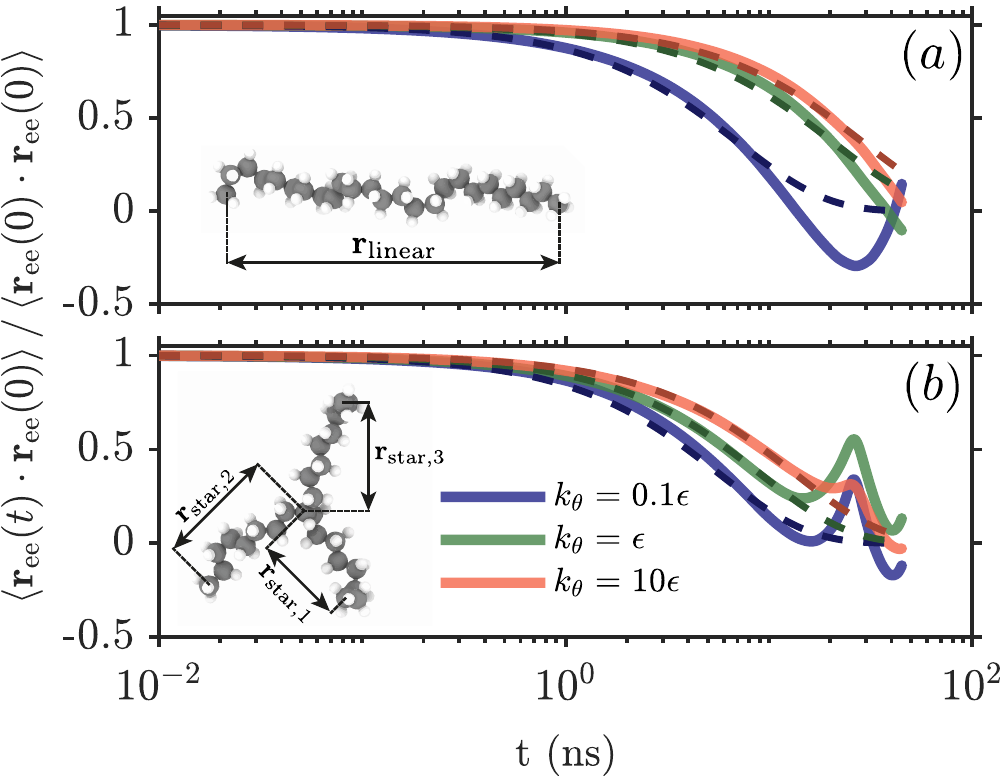}
    \caption{Calculation of the relaxation times of the linear and star polymers for varying chain stiffness. Normalized time-autocorrelation function vs. time for (a) linear and (b) star polymers. The dashed lines denote the exponential fit associated with TACF.}
    \label{fig:fig1}
\end{figure}
%LB IN FIGURE 1 the y-axis should be r_ee not r: The reason why I did not use r_ee but r was we didn't have a proper end-to-end distance expression for the branched polymers and I showed what r meant in the small illustrations of the molecules. But since it defines the end-to-end distance, I can surely add _ee. (ABY)

The relaxation time of a polymer chain, $\lambda$, can be related to the timescale, in which its deformed state returns to its undeformed configuration. Thus, the competition between $\lambda$ and the shear time scale can be used to define the deviations from equilibrium~\cite{Erbas2015,Semenov:1995gg}, requiring us to estimate $\lambda$. For this reason, we refer to the time autocorrelation function (TACF) for the Rouse model, which is valid for melt, and use the chain end-to-end vector $\mathbf{r_{ee}}$ %CHECK IF IT IS CORRECT LB: Formula is correct. (ABY) 
\begin{equation}
    \frac{\langle\mathbf{r_{ee}}(t) \cdot \mathbf{r_{ee}}(0)\rangle}{n b^2}=\sum_{\mathrm{p}=1,3, \ldots} \frac{8}{\mathrm{p}^2 \pi^2} \exp \left(-\frac{t \mathrm{p}^2}{\tau_1}\right),
\end{equation}
where $n$ is the number of carbon atoms in a chain, $b^2$ is the mean-square distance between adjacent beads (hence $nb^2$ is the equilibrium mean-square end-to-end distance of chains), $\tau_p$ is the spectrum of the relaxation times for the corresponding mode $p$. A visual definition of $\mathbf{r_{ee}}$ is given in the inset of Fig. \hyperref[fig:fig1]{1(a)}.
The relaxation dynamics are dominated by the slowest mode ($\mathrm{p}=1$)~\cite{doxastakis_chain_2003, kalathi_rouse_2014}, allowing us to use $\lambda \approx \tau_1$. Note that  the definition of $\mathbf{r_{ee}}$ is not straightforward for branched polymers owing to the presence of more than two free ends \cite{hu_relaxation_2019}; thus, we modify the $\mathbf{r_{ee}}$ definition for star polymers by considering the branching point to the corner vectors and track the mean-square end-to-end vectors over time (Fig. \hyperref[fig:fig1]{1(b)}, inset). 

To obtain $\lambda$ for linear and star polymers from our simulations, we calculate the equilibrium (i.e., no shear) TACF for three chain rigidities by using our trajectories: $k_{\theta}=0.1\epsilon$, $k_{\theta}=\epsilon$ and $k_{\theta}=10\epsilon$ by fitting the correlation function to $ {\langle\mathbf{r_{ee}}(t) \cdot \mathbf{r_{ee}}(0)\rangle}/{\langle\mathbf{r_{ee}}(0) \cdot \mathbf{r_{ee}}(0)\rangle} \propto \exp \left ( -t/\lambda \right )$ (Fig. \hyperref[fig:fig1]{1}). We obtain the Rouse times as $\lambda_{\mathrm{linear}}^{(0.1\epsilon)} = 7.07$ ns, $\lambda_{\mathrm{linear}}^{(1\epsilon)} = 20.8$ ns, $\lambda_{\mathrm{linear}}^{(10\epsilon)} = 29.14$ ns, for linear polymers; $\lambda_{\mathrm{star}}^{(0.1\epsilon)} = 5.73$ ns, $\lambda_{\mathrm{star}}^{(1\epsilon)} = 9.10$ ns, $\lambda_{\mathrm{star}}^{(10\epsilon)} = 13.53$ ns, for star polymers. As the chain flexibility increases $\left ( \mathrm{i.e.,~} k_{\theta} \rightarrow 0 \right )$, the chain needs less time to recover, asymptotically leading to $\lambda \rightarrow 0$ in the case of monomers \cite{rudisill_non-equilibrium_1993}. This relation between $\lambda$ and chain rigidity $\left (\mathrm{i.e.,~} \lambda^{(10\epsilon)}>\lambda^{(1\epsilon)}>\lambda^{(0.1\epsilon)} \right)$ is quantitatively consistent with the previous results~\cite{kalathi_rouse_2014, moreira_direct_2015}, and so does the relation between $\lambda$ and the polymer topology $\left (\mathrm{i.e.,~} \lambda_\mathrm{linear}>\lambda_\mathrm{star} \right)$~\cite{jabbarzadeh_effect_2002}. Moreover, stiffening the chains increases the Kuhn size, which results in varying the chain rigidity to scale with the Kuhn sizes \cite{everaers_rheology_2004, kirby_nodate, ding_comment_2004,everaers_rheology_2004}.

\begin{figure}[b]
    \centering
    \includegraphics[width=0.475\textwidth]{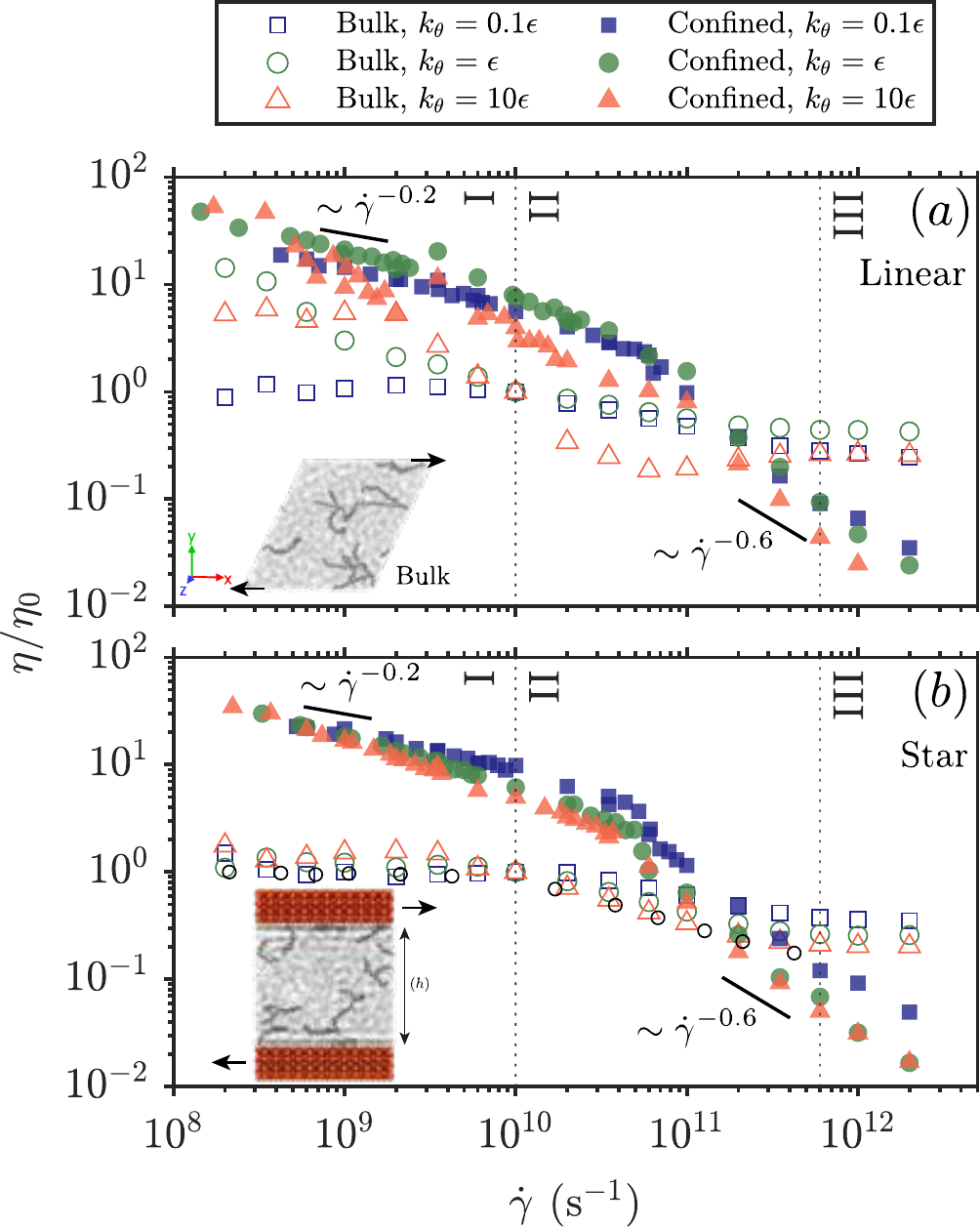}
    \caption{Viscosity normalized with respect to the Newtonian plateau for the bulk (empty markers) and the confined (filled markers) configurations for the linear (a) and the star polymers (b). The black circles denote the viscosity values obtained by McCabe et al. \cite{mccabe_examining_2001} for unaltered star polymer ($k_{\theta} = \epsilon$). Error bars are of the size of markers.}
    %\caption{Normalized viscosity with respect to Newtonian plateau. Bulk and confinement viscosity of linear polymer (a) and star polymer (b). $\eta_{0,\mathrm{linear}}^{0.1\epsilon} = 0\;\mathrm{cP}$, $\eta_{0,\mathrm{linear}}^{\epsilon} = 0\;\mathrm{cP}$, $\eta_{0,\mathrm{linear}}^{10\epsilon} = 0\;\mathrm{cP}$; $\eta_{0,\mathrm{star}}^{0.1\epsilon} = 0\;\mathrm{cP}$, $\eta_{0,\mathrm{star}}^{\epsilon} = 0\;\mathrm{cP}$, $\eta_{0,\mathrm{star}}^{10\epsilon} = 0\;\mathrm{cP}$. Black circles denote the viscosity values obtained by McCabe et al. \cite{mccabe_examining_2001}}.
    \label{fig:fig2}
\end{figure}

Next, we compare the steady-state rheology of the confined polymer melts to their bulk state under planar Couette flow to explore the confinement effects. We run shear simulations of (i) the bulk melts by deforming the simulation box (Fig. \hyperref[fig:fig2]{2(a)}, inset) and (ii) the confined melts by moving the top and bottom surfaces (i.e., fixed and rigid hematite, $\alpha-\mathrm{Fe_{2} O_{3}}$, layers) at constant velocity in the $\pm x$-directions (Fig. \hyperref[fig:fig2]{2(b)}, inset). We set the confinement level to $h = 5.57$ nm (for $N=216$) to ensure that the pressure over the walls is the same as the hydrostatic pressure in bulk (i.e., $p = 1$ atm). Thus, any change in the (microscopic) stress tensor purely represents the flow characteristics. 

From the simulation trajectories, the microscopic stress tensor $\bm{\upsigma}$ can be calculated using the virial theorem \cite{itoh_enhanced_2018, yang_calculation_2014, todd_nonequilibrium_2017}. Consequently, the ensemble-averaged total stress tensor, composed of the kinetic and virial contributions, is prone to thermal fluctuations \cite{yang_generalized_2012}. Therefore, we compute the total stress tensor of the melt in bulk and confined configurations and consider the time-averaged steady-state stress to eliminate thermal fluctuations. We define the macroscopic equivalent formulation \cite{Engineering_rheology}
\begin{equation}
\bm{\upsigma} \equiv  \underbrace{-p\mathbf{I} + 2\eta\left ( \bm{\dot{\upgamma}} \right )   \bm{\dot{\upgamma}}}_{\mathrm{GNF}} + \bm{\upsigma}^{\left ( \mathrm{VE} \right )},   
\label{eq:EngRheo}
\end{equation}
where $p$ is the mechanical pressure of the melt, $p = - \frac{1}{3} \left ( {\upsigma_\mathrm{{xx}}} + {\upsigma_\mathrm{{yy}}} + {\upsigma_\mathrm{{zz}}} \right )$, 
$\bm{\dot{\upgamma}}$ is the strain rate tensor with the only nonzero components, namely the shear rate, $\dot{\upgamma} = \dot{\upgamma}_\mathrm{{xy}} = \dot{\upgamma}_\mathrm{{yx}}$, and the extra-stress component for the viscoelastic effects $\bm{\upsigma}^{\left ( \mathrm{VE} \right )}$. As $\bm{\upsigma}^{\left ( \mathrm{VE} \right )} \rightarrow 0$, Eq.~\ref{eq:EngRheo} reduces to the GNF stress tensor, which introduces the shear-dependency to the Newtonian viscosity $\eta_0$ \cite{tseng_revisitation_2020, deville_mathematical_2012, Engineering_rheology, ahmed2022modified}. 

To characterize the shear-dependent viscosity of our melts, we calculate the normalized  viscosity, $\eta/\eta_0$, where the shear viscosity is defined as $\eta \equiv \upsigma_{xy} / \dot{ \upgamma}$, and  the Newtonian viscosity is $\eta_0$ \cite{bair_comparison_2002}. The observed behaviour of viscosity for bulk and confinement shows the expected shear-thinning behaviour (for linear: Fig. \hyperref[fig:fig2]{2(a)}, for star: Fig. \hyperref[fig:fig2]{2(b)}) \cite{mccabe_examining_2001}. In Fig. \hyperref[fig:fig2]{2}, three regions are labelled to indicate the various shear responses; \Romannum{1}, \Romannum{2}, and \Romannum{3} indicate the \nth{1} Newtonian plateau, the shear-thinning regime, and the \nth{2} Newtonian plateau for bulk polymers, respectively. 

The linear polymers in bulk tend to form crystalline structures at the Newtonian plateau for $k_{\theta} = \epsilon$ and $k_{\theta} = 10\epsilon$ cases. This leads to a higher viscosity than the zero-shear viscosity \cite{qu_flow-directed_2016, nafar_sefiddashti_flow-induced_2020}, independent of the size of our simulation boxes (see Supplemental Material, Fig. S1). The flow-induced crystallization starts to vanish as the shear rate increases \cite{boutaous_thermally_2010, graham_molecular_2011, jabbarzadeh_flow-induced_2010}. Notably, we do not observe crystallization for strongly flexible chains (i.e., $k_{\theta} = 0.1\epsilon$), which is a common simulation scheme for coarse-grained models ~\cite{grest_polymer_1996,everaers2004rheology}.
In the confined geometry, the viscosity of both polymer topologies is higher than their bulk states and monotonically decreases with the shear rate (Fig. \hyperref[fig:fig2]{2}) \cite{itoh_enhanced_2018}. This decrease follows the power law $\eta \sim \dot{\upgamma}^{-0.2}$ until shear bands start to develop in region \Romannum{2}. In the transition regime, the Couette velocity profile starts to become centrally localized in the middle of the channel, disrupting the smoothness of the velocity profile \cite{ewen_contributions_2021, ewen_effect_2017, nimura_viscoelasticity-induced_2022}. With increasing shear rate, a discontinuity in the velocity profile develops, resulting in plug slip behaviour, and the decrease in viscosity follows the power law $\eta \sim \dot{\upgamma}^{-0.6}$ (see Supplemental Material, Figs. S2 and S3 for velocity profiles). As the flexibility of the chains increases, the shear-thinning behaviour becomes stronger. Notably, if we replace our chains with monomers (i.e., no chains) and run separate simulations, no additional resistance due to the confinement appears (see Supplemental Material, Fig. S4), highlighting the role of chain connectivity.

\begin{figure*}[tp]
    \centering
    \includegraphics[width=0.95\textwidth]{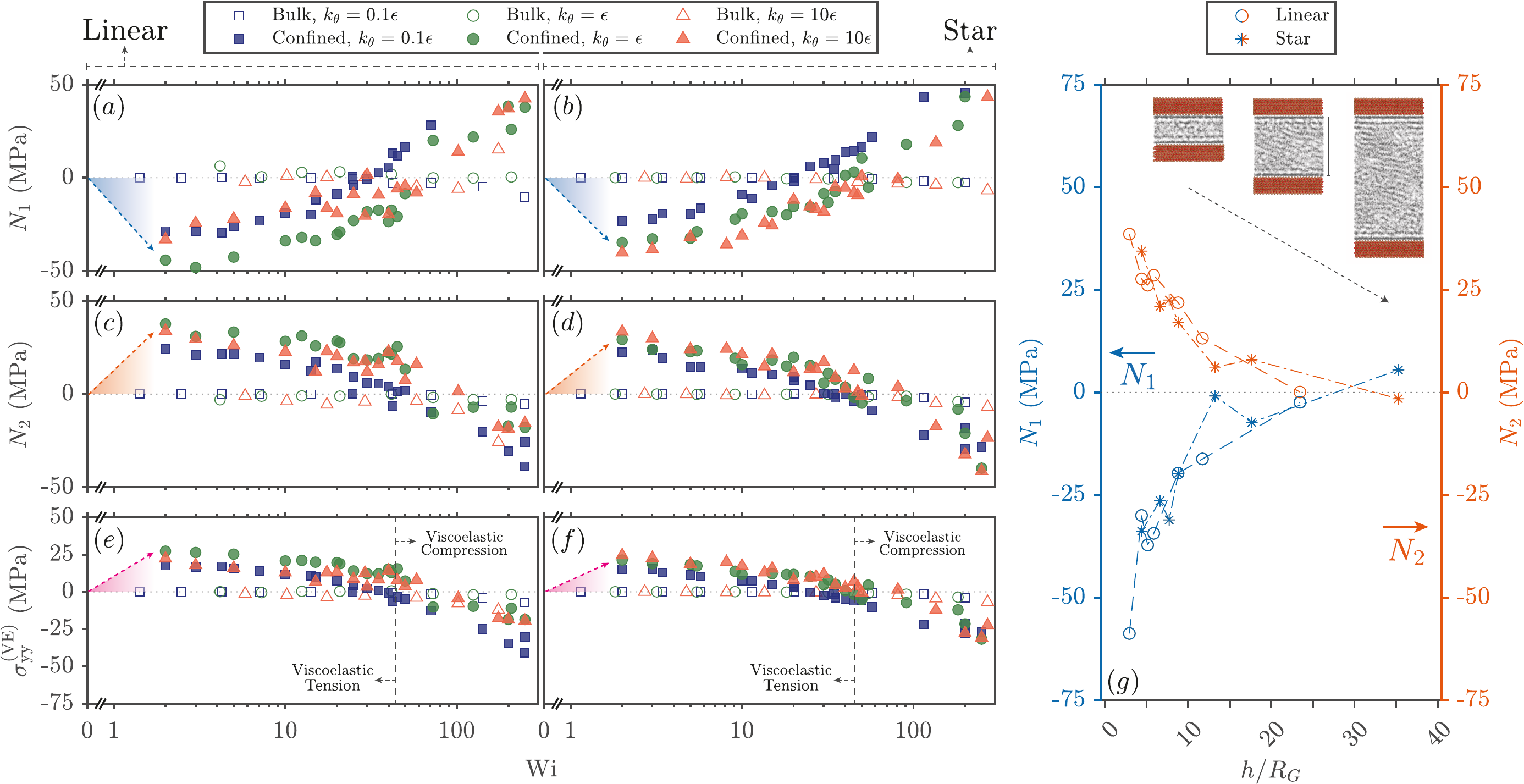}
    \caption{Normal stress differences $N_1,\;N_2$ and viscoelastic stress component opposing the streamwise direction ${\upsigma}^{\left ( \mathrm{VE} \right)}_{yy}$. $N_1$ for linear (a) and star (b) polymers). $N_2$ for linear (c) and star (d) polymers. ${\upsigma}^{\left ( \mathrm{VE} \right)}_{yy}$ for linear (e) and star (f) polymers, highlighting the regions of viscoelastic tension and compression applied to the walls. (g) $N_1$ and $N_2$ for the ratio of film thickness $h$ to the radius of gyration $R_G$ at $Wi$ = 10 (see Supplemental Material, Fig. S5 for $R_G$). Note that the horizontal axes (a-d) are interrupted. Error bars are of the size of markers.}
    \label{fig:fig3}
\end{figure*}

To quantify the viscoelastic contributions to the total rheological response, we turn our attention back to the stress tensor $\bm{\upsigma}$ (Eq. \ref{eq:EngRheo}) and focus on the viscoelastic part of the stress tensor, $\bm{\upsigma}^{\left ( \mathrm{VE} \right )}$. According to Larson \cite{larson_2019}, viscoelasticity originates from unequal normal stress components, characterized by normal stress differences, with notable exceptions of the suspensions of non-Brownian non-spherical particles and porous polymer gels \cite{maklad_review_2021, osuji_highly_2008, montesi_vorticity_2004, de_cagny_porosity_2016}. Hence, the viscoelastic stress is a function of the normal stress differences, $\bm{\upsigma}^{\left ( \mathrm{VE} \right ) }\left(N_1, N_2 \right)$ \cite{Engineering_rheology}. For concentrated polymeric liquids,  the first and second normal stress differences ($N_1 = \upsigma_\mathrm{xx} - \upsigma_\mathrm{yy}$ and $N_2 = \upsigma_\mathrm{yy} - \upsigma_\mathrm{zz}$) arise due to the interactions between the chains trapped near the surfaces and  the micro-structure dynamics, respectively \cite{lin_normal_2014}. Normal-stress differences govern the viscoelasticity in the scope of polymer melts, meaning that shear-thinning fluids are not necessarily viscoelastic as well \cite{larson_2019, maklad_review_2021}.

In Fig. \hyperref[fig:fig3]{3}, we explore the normal stress differences and the viscoelastic stress component as a function of the Weissenberg number ($Wi=\dot{\gamma}\lambda$) and the film thickness $h$. The Weissenberg number essentially characterizes the degree of non-linearity in the deformation behaviour of viscoelastic materials. When $Wi<<1$, the fluid exhibits characteristics similar to a Newtonian fluid. However, the fluid displays viscoelastic effects for $Wi\geq O(1)$ as normal forces become more dominant than shear forces, causing the polymers to stretch and store elastic energy \cite{morozov_introductory_2007}. Noting that $N_1 = N_2 = {\upsigma}^{\left ( \mathrm{VE} \right )}_\mathrm{yy} = 0$ for $Wi = 0$, we observe an unexpected behavior for the normal stress differences; $N_1<0$ and $N_2>0$ leading to viscoelastic tension for $Wi<50$. Within the practical limits of our MD simulations, we determine that ${\upsigma}^{\left ( \mathrm{VE} \right )}_\mathrm{yy}$ increases up to $Wi\approx$ 3 and decreases afterward (Fig. \hyperref[fig:fig3]{3(a-f)}, shown with arrows). Approximately after $Wi = 100$, both the normal stress differences change sign, $N_1>0$ and $N_2<0$, also yielding ${\upsigma}^{\left ( \mathrm{VE} \right )}_\mathrm{yy} < 0$ (i.e., viscoelastic compression). When confined within the same $Wi$ interval, the same behaviour is present for both linear and star polymer topologies. On the other hand, for bulk systems, we observe $N_1 = N_2 = {\upsigma}^{\left ( \mathrm{VE} \right )}_\mathrm{yy} \approx 0$, suggesting that the storage of the elastic energy depends on the restriction of movement of chains once confined. 

To explore the change in $N_1$ and $N_2$ as a function of $h$, we change the film thickness while keeping the density constant (i.e., for each thickness $h$, $N/h$ is preserved by varying $N$). In this way, we show how the normal stress differences increase in magnitude as the film thickness decreases (Fig. \hyperref[fig:fig3]{3(g)}).
% Within the viscoelastic boundary layer, the viscoelastic tension may emerge and increase the load-carrying capacity. 
%LB this sentence may need to be moved while commenting fig. 4 since we did not introduce yet the viscoelastic boundary layer: Makes sense, done (ABY)
As $h$ increases, ideally approaching the bulk system, the normal stress differences decrease substantially in magnitude and even change sign while retaining the same ratio $N_1/N_2$, resulting in the well-known behavior of polymeric liquids with $N_2<0$ \cite{dlugogorski_viscometric_1993, thien_new_1977, rudisill_non-equilibrium_1993, jabbarzadeh_effect_2002, maklad_review_2021,shaw_introduction_nodate, larson_2019}. As achieving smaller film thickness becomes experimentally harder to achieve, this hints at why negative $N_1$, positive $N_2$, and hence viscoelastic tension are not usually observed via experiments while having large magnitudes \cite{shaw_introduction_nodate,schweizer_measurement_2002}. As the normal stress differences, and hence the viscoelastic stress component, become smaller for larger $h$, constitutive models, such as GNF, show less deviation from the real microscopic behavior (Figs. \hyperref[fig:fig3]{3(a-f)}).

We relate the unexpected behavior of the normal stress differences to the alignment of the bonds constituting the chains since we do not observe significant changes in the chain end-to-end distance (see Supplemental Material, Fig. S6 for $R_{G,x}$ and $R_{G,y}$ behavior near - and away from - the surfaces)~\cite{granick_polymers_1999, kirk_surface_2018, cho_molecular_2017,israelachvili_intermolecular_nodate,tsige_all-atom_2008}. For this purpose, we calculate the bond-orientation tensor, where we represent each bond independently, with the following formulation \cite{yasuda_synchronized_2014,galvani_cunha_probing_2022}
\begin{equation}
    Q_{\alpha \beta}= \frac{1}{N} \sum_{N}  \frac{1}{N_b-1} \sum_{j=1}^{N_b-1} \frac{{b_{j \alpha} \; b_{j \beta}}}{\left \| b \right \|^2},
\end{equation}
where $N$ is the number of chains, $N_b$ is the number of carbon atoms in a polymer chain, $b_j$ for $1 \leq j \leq N_b - 1$ is the bond vector between consecutive carbon atoms in the same chain, and $\left \| b \right \|$ is the Euclidean length of the bond. To capture the near-surface dynamics, we refer to the $yy$-component of the bond-orientation tensor (i.e., $Q_{yy}$), representing the alignment of bonds against the flow. 

\begin{figure}[!t]
    \centering
    \includegraphics[width=0.475\textwidth]{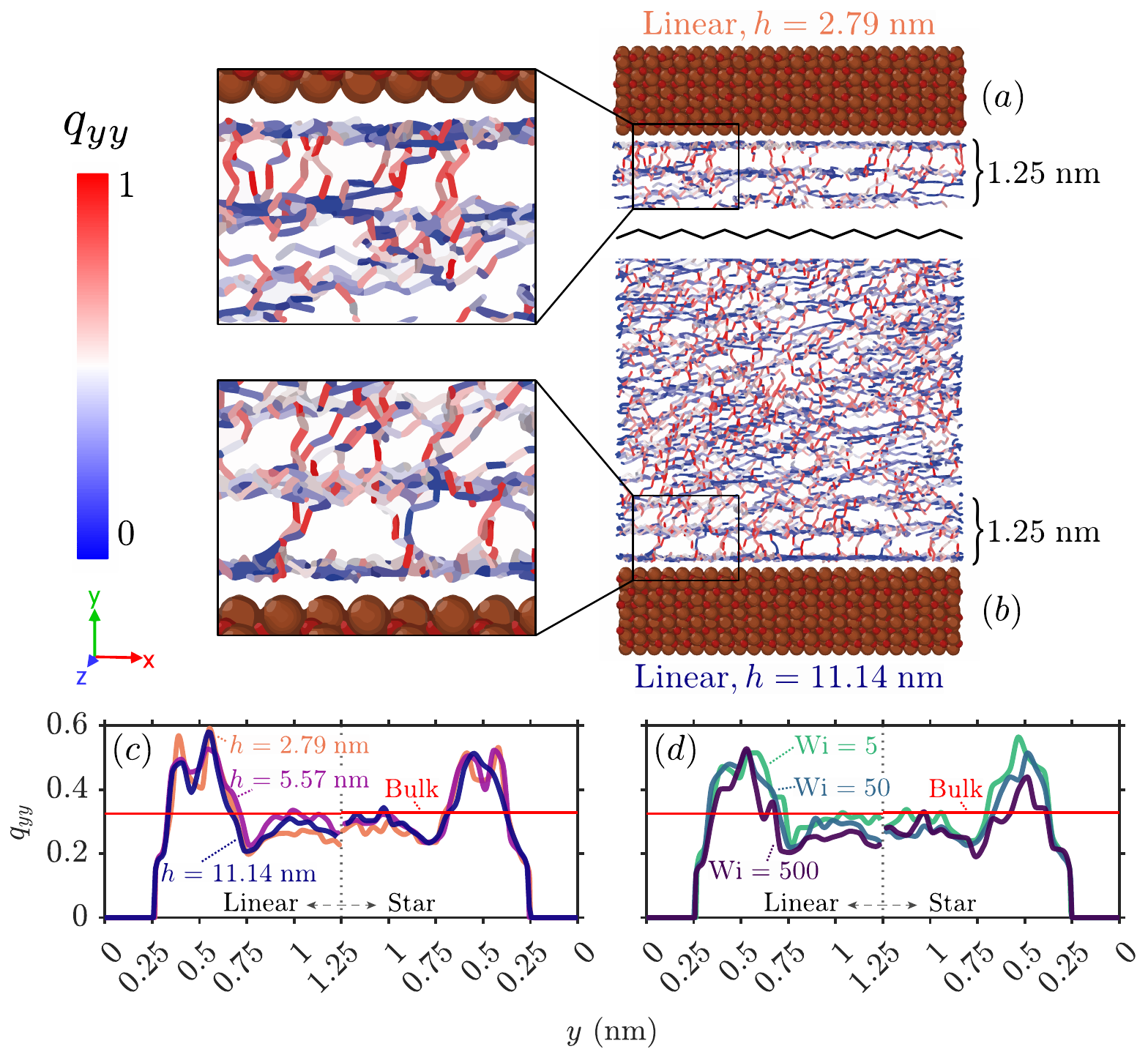}
    \caption{Local bond orientation, $q_{yy}$. Snapshots of the spatial resolution of $q_{yy}$, for $h=2.79$ nm (a) and $h=11.14$ nm at $Wi = 10$ for linear polymers. Spatial variation of $q_{yy}$ for different $h$ for linear (left) and star polymers (right) (c)  and for varying $Wi$ for linear (left) and star polymers (right) (d).}
    % \caption{Local bond orientation, $q_{yy}$. Snapshots of spatial resolution of $q_{yy}$ of the half simulation box, (c, d) for varying $h$ at $\mathrm{Wi} = 10$; (e, f) for varying Wi for $h = 5.57$ nm. Spatial change in $q_{yy}$ for varying film thickness for linear (left) and star polymer (right) (e)  and for varying Wi for linear (left) and star polymer (right) (f). }
    %\caption{Local bond orientation, $q_{yy}$. Change in bond orientation for varying film thickness for linear polymer (left) and star polymer (right) (a).  Change in bond orientation for varying Wi for linear polymer (left) and star polymer (right) (b). Snapshots illustrating near-surface $q_yy$ for $h$/2, (c, d) are for $\mathrm{Wi} = 10$ while $h$ varies, and (e, f) are for $h = 5.57$ nm while Wi varies.}
    \label{fig:fig4}
\end{figure}

To characterize the spatial variation in $Q_{yy}$, we consider $Q_{yy} = \int^h_0 q_{yy}\left ( y\right )dy$ with the discretization $Q_{yy} \approx \sum^h_0 q_{yy}\left ( y\right )\Delta y$, where $\Delta y \approx 0.5 \si{\angstrom}$. %\hyperref[fig:fig4]{Fig. 4}
With this, we exploit the \textit{local} bond orientation $q_{yy}$ among confined channels in Fig. \hyperref[fig:fig4]{4}, where the representative snapshots color-codes the bond orientation such that the bond aligned vertically to the flow are red and those horizontal are given in blue. A visual inspection reveals a strongly absorbed polymer layer between the surface and the bulk chains. This \textit{viscoelastic} layer rather resembles a polymer brush with surface-grafted chains ~\cite{Erbas2015}. Notably, throughout our simulations, adsorbed chains do not exchange with the bulk chains (see the movie S7 in the Supplemental Material). When the surfaces are separated from each other, although the near-surface peak of $q_{yy}$ is preserved, the bonds, overall, become more aligned with the flow in the $x$-direction throughout the channel, leading to a decrease in $q_{yy}$ (Fig. \hyperref[fig:fig4]{4(c)}). At relatively fast shear rates (e.g., $Wi>50$), the bonds become more aligned in  the flow direction, leading to a decrease in $q_{yy}$. At lower shear rates (e.g., $Wi<50$), the chains have more time to reorient to their entropically favoured states, creating additional viscoelastic force acting on the surfaces (Fig. \hyperref[fig:fig4]{4(d)}) \cite{arnold_unexpected_2007, oconnor_relating_2018, xie_chemical_2019}. Consequently, at lower $Wi$, the load-carrying capacity of the melt increases drastically, resulting in stresses orders of magnitude larger than in bulk (${\upsigma}^{\left ( \mathrm{VE} \right )}_\mathrm{yy} \sim 25$ MPa vs. $p \sim 0.1$ MPa, see Fig. \hyperref[fig:fig3]{3(e,f)}). Importantly, the thickness of the polymer-adsorbed layer is independent of $h$, and thus, the contribution of this layer to the overall viscoelastic response decreases as $h/R_G \rightarrow \infty$.

%We relate the increase in the magnitude of $N_1$ and $N_2$ as $h$ decreases, given in Fig. \hyperref[fig:fig3]{3(g)}, \textbf{to the emergence of bonds between the melt coating the surface and the remaining bulk part }\cite{granick_polymers_1999}. These bonds bridging two regions are observed to remain static (i.e., no other bridging bonds emerge during the flow). When $h$ increases, although the near-surface peak is preserved, the bonds, overall, become more aligned with the flow in the $x$-direction throughout the channel, leading to a decrease in $q_{yy}$ (Fig. \hyperref[fig:fig4]{4(c)}). Similarly, we relate the behaviour of $N_1$ and $N_2$, given in Fig. \hyperref[fig:fig3]{3(a-f)}, to the alignment of bonds with the flow. Within the viscoelastic boundary layer, viscoelastic tension may emerge, increasing the load-carrying capacity. Consequently, at lower $Wi$, the load-carrying capacity of the melt increase drastically, resulting in $\sim 250$ times larger forces compared to unaltered star polymers mediated at $p=1$ atm.

Our results demonstrate unexpected, sign-reversed normal stresses, namely $N_1$ and $N_2$, together with the emergence of a viscoelastic tension and compression as the confinement dimension approaches that of the confined molecules. We observe a peculiar orientation of the near-surface chains. Our simulations suggest that the chains near the surface can be responsible for the elastic force exerted on the surface walls, and this effect can decay as the chains become more aligned with the flow at higher shear rates (i.e., $Wi\gg1$). Excess tension emerges within a viscoelastic boundary layer, in which chains are vertically aligned, resulting in an increase in the load-carrying capacity of the confined melt. We observe smaller (larger) stresses for more flexible (rigid) chains for the two chain topologies we consider here, namely linear and star polymers. Notably, the stress components show weak sensitivity to the persistence length. We further reveal the dependence of the relaxation time on the chain rigidity, such that, by varying the chain rigidity, we modify the contribution of the elastic forces without altering the molecular structure \cite{everaers_rheology_2004}. Our study is limited to the polymer melts of linear and star polymers confined with rigid iron-oxide walls. Introducing the flexibility of the walls should reduce the observed effect, and slip and adsorption may depend on the polymer surface type \cite{bernardi_thermostating_2010, ta_thin_2015, gattinoni_adsorption_2018, kanhaiya_accurate_2021,napolitano_irreversible_2020,krim_friction_2012, apostolo_molecular_2019}. Furthermore, it is yet to be explored whether this effect is typical for polymer melts, observable in more complex polymer liquids under higher pressures or different surface roughness conditions \cite{savio_model_2012, albina_coarse-grained_2020, halverson_rheology_2012}. 

The authors would like to acknowledge the The Scientific and Technological Research Council of Turkey (TUBITAK) for supporting this work under the project 221M576. The numerical calculations reported in this paper were partially performed at the TÜBİTAK ULAKBİM, High Performance and Grid Computing Center (TRUBA resources).

%REF. 9 seems wrong! check LB: All the references are auto-generated by Zotero. I will find the correct reference style for PRLs and change every single one accordingly. (ABY)

%% DETAILS TO MENTION

% N1 increases quadratically with strain rate \cite{morozov_introductory_2007}

% Lubrication is related to RoG before - while mentioning FIG4 \cite{liu_lubricant_2017, korolkovas_polymer_2017}

% Positive YY is observed near-surface \cite{morciano_nonequilibrium_2017}

\bibliographystyle{h-physrev}

\bibliography{Bib1}
\end{document}

%\preprint{APS/123-QED}
\title{Supplemental Material: Molecular Rheology of Nanoconfined Polymer Melts}
\author{Ahmet Burak Yıldırım}
\affiliation{Department of Mechanical Engineering, Bilkent University, 06800 Ankara, Turkey}

\author{Aykut Erbaş}
\affiliation{UNAM - National Nanotechnology Research Center and Institute of Materials Science \& Nanotechnology, Bilkent University, 06800 Ankara, Turkey}
\affiliation{Institute of Physics, University of Silesia, Katowice, Poland}

\author{Luca Biancofiore}
\affiliation{Department of Mechanical Engineering, Bilkent University, 06800 Ankara, Turkey}
\affiliation{UNAM - National Nanotechnology Research Center and Institute of Materials Science \& Nanotechnology, Bilkent University, 06800 Ankara, Turkey}

\date{\today}% It is always \today, today,

\maketitle
\section{Simulation Details} 
The work includes two types of simulations: (i) diffusion to determine the relaxation times and (ii) shear to determine the viscoelastic response. For both types, the interactions between all atoms are modelled by the standard 12/6 Lennard-Jones potential using D\textsc{reiding} force field \cite{mayo_dreiding_1990, das_solvation_1996}, with the cutoff radius of $r_c = 9.85$ $\mathrm{A}^\circ$ and the time-integration is performed for the timestep of $\tau = 1$ fs. 

For (i), each simulation box contains $N=216$ charge-neutral linear or star $\mathrm{C_{25} H_{52}}$ polymers equilibrated at $T = 372$~K at a pressure at $p = 1$ atm ~\cite{mccabe_examining_2001}. The resultant melt density of $\rho = 0.63$ $\mathrm{g/cm^3}$ is expected as D\textsc{reiding} potential is known to underestimate the density of the polymer melts while giving accurate viscosity estimations \cite{ewen_comparison_2016}. Preserving the density does not guarantee that hydrostatic pressures are also identical between the systems; the hydrostatic pressures are also prone to variations in the confinement level, chain topology, and rigidity. However, since we are mainly interested in acquiring identical geometric confinements and densities for all of our melt systems, we disregard the variations in hydrostatic pressure between the systems \cite{kontopoulou_basic_2011} but rather track the change in the flow-dependent (i.e., dynamic) components of the stress tensor. This helps us to construct $N_1$, $N_2$, $\bm{\upsigma}^{\left ( \mathrm{VE} \right )}$ independently of the hydrostatic pressure to study the rheological response. 
We track the mean-square end-to-end vectors for $t=10^8~\tau$ under NVE dynamics regulated at $T=372$~K using the Langevin thermostat with a damping coefficient of $100~\tau$. We later fit the correlation function following $ {\langle\mathbf{r_{ee}}(t) \cdot \mathbf{r_{ee}}(0)\rangle}/{\langle\mathbf{r_{ee}}(0) \cdot \mathbf{r_{ee}}(0)\rangle} = A\exp \left ( -t/\lambda \right )$, with constant $A$ ($\approx 1$ for each fit).

For (ii), the simulation boxes contain varying $N$ polymers, depending on the box size or confinement level, while keeping the density $N/V$ constant for the volume $V$. For the bulk system, we run shear simulations by deforming the cubic simulation box ($h = 5.82$ nm, resulting in $p = 1$ atm for unaltered star polymer) under constant volume and temperature by solving S\textsc{llod} equations of motion \cite{raghavan_shear-thinning_2017} at $T=372$~K with a damping coefficient of $100~\tau$. For the confined melts, we run shear simulations under NVE dynamics at $T=372$ K using the Langevin thermostat ($N = 216$ for box dimensions: 6.35 nm $\times$ 5.57 nm $\times$ 6.97 nm) with a damping coefficient of $10~\tau$, which is achieved by moving the top and bottom surfaces at constant velocity in the x-direction \cite{morciano_nonequilibrium_2017}. While investigating the confinement effect on $N_1$ and $N_2$, we changed the number of polymers $N$, accordingly (i.e., $N=216, h = 5.57$ nm). The thermostat is only applied to the non-sheared directions to preserve the increase in velocity due to shear correctly. The fixed and rigid wall configuration consists of five hematite $ \left ( \alpha-\mathrm{Fe_{2} O_{3}} \right ) $ layers on the top and bottom surfaces \cite{ewen_nonequilibrium_2016}, which  results in a wall thickness of 1.31 nm to ensure a negligible wall deformation compared to thin and deformable walls. Unlike soft walls with variable film thickness, rigid walls are known to experience higher load variations, making them a better choice for observing the magnified confinement effects \cite{israelachvili_intermolecular_nodate}. A Couette flow is simulated for $10^6-2\times10^7\tau$, ensuring that a typical polymer chain passes through the periodic boundaries at least three times, eliminating the transient response. The damping coefficients are selected to ensure that the temperature remains $\sim 372\pm 0.1$~K throughout the simulations. 

Note that the volume of the bulk and confined systems differ due to wall interactions; hence $p = 1$ atm is preserved for unaltered star polymers while determining the volume of both linear and star polymer systems. This enables us to validate some of the results for star polymers with Ref.~\cite{mccabe_examining_2001}.

\section{Crystallization}
\begin{figure}[H]
    \centering
    \includegraphics[width=0.99\textwidth]{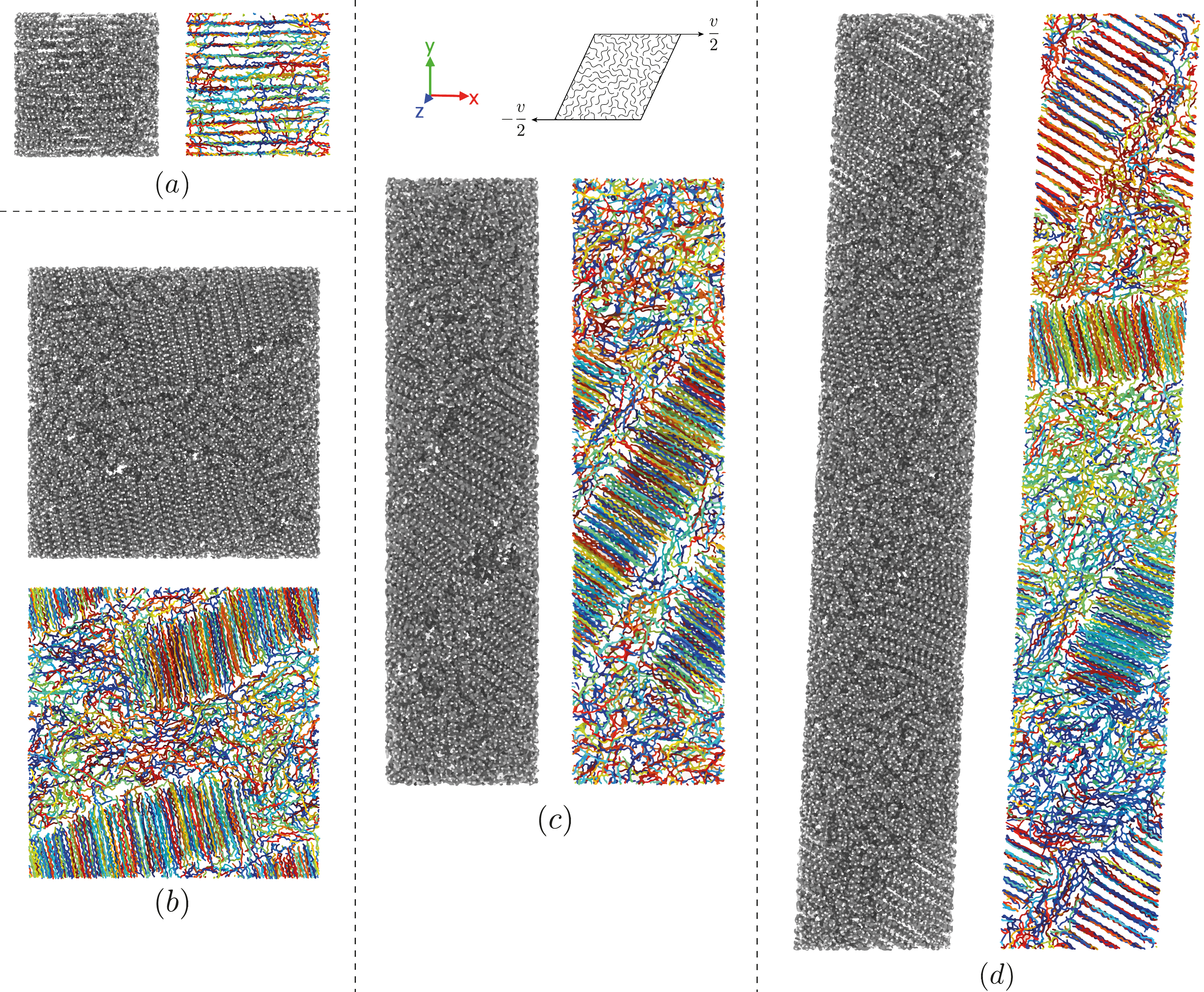}
    \caption{Snapshots of the crystalline structure formation of unaltered linear polymer melt ($k_{\theta} = \epsilon$) under shear for different simulation box sizes. The shear rate is $\dot{\gamma}=6\times10^8 s^{-1}$, corresponding to $Wi = 12.48$. Size of the simulation boxes for (a) is $(h,h,h)$, (b) is $(2h,2h,h)$, for (c) is $(h,4h,h)$, and (d) is $(h,6h,h)$ where $h \approx 5.82 \si{\angstrom}$ to retain the identical density of $\rho = 0.63$ $\mathrm{g/cm^3}$. Under identical settings, star polymer melts do not exhibit crystallization. For each simulation box size, molecular and bond configurations are illustrated separately, where the coloring of the bonds is random.} 
    \label{fig:fig1}
\end{figure}

\section{Slip}

\begin{figure}[H]
    \centering
    \includegraphics[width=0.99\textwidth]{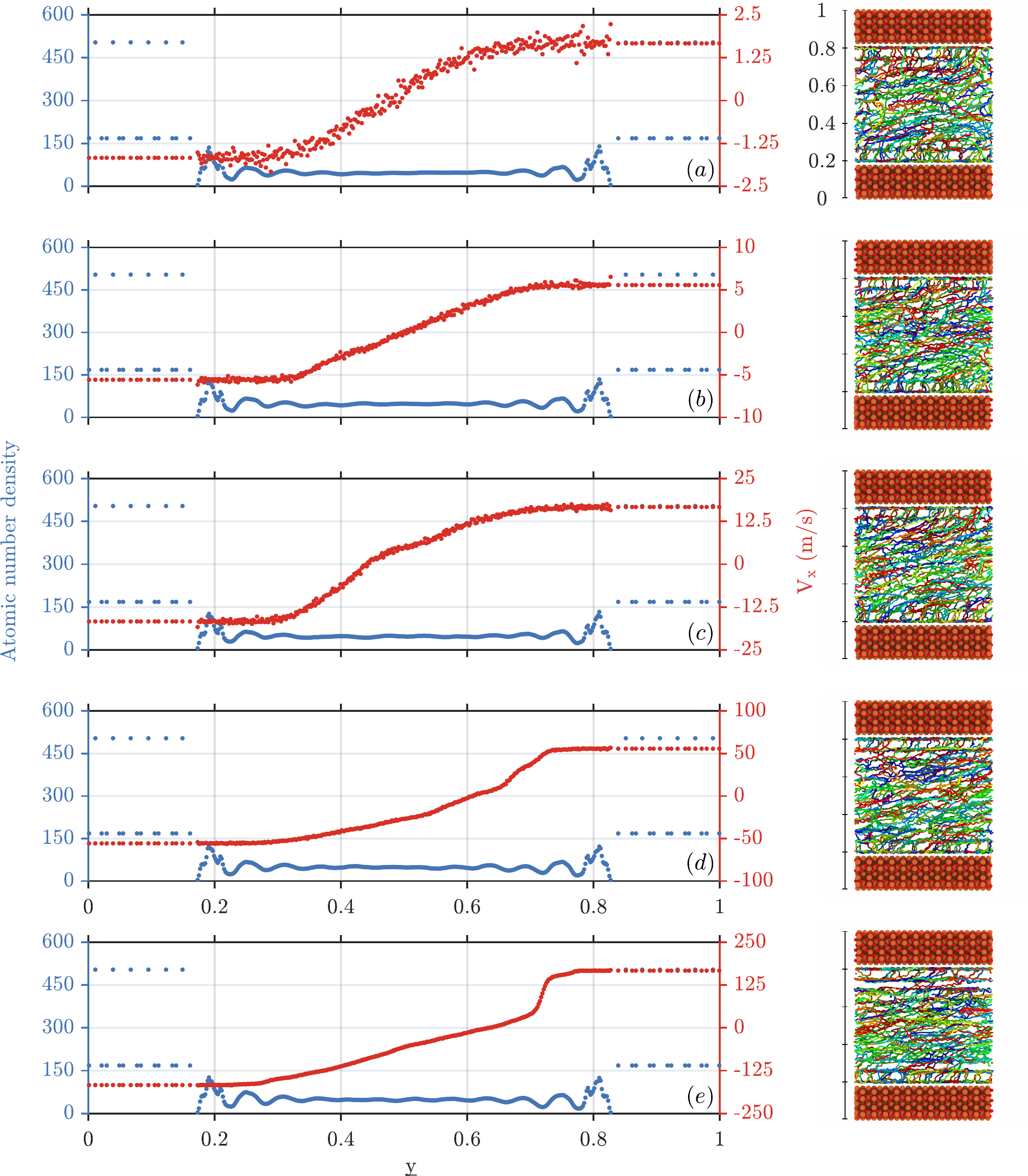}
    \caption{The steady-state spatial configuration and the velocity distribution of the atoms within the shear of confined linear polymer melts for $h = 5.57 \si{\angstrom}$. The shear rate is (a) $\dot{\gamma}=6\times10^8 s^{-1}$, (b) $\dot{\gamma}=6\times10^9 s^{-1}$, (c) $\dot{\gamma}=6\times10^{10} s^{-1}$, (d) $\dot{\gamma}=6\times10^{11} s^{-1}$, and (e) $\dot{\gamma}=6\times10^{12} s^{-1}$ , as a function of the normalized vertical component $\underline{y}$. The corresponding Weissenberg numbers are (a) $Wi = 12.48$, (b) $Wi = 124.8$, (c) $Wi = 1248$, (d) $Wi = 12480$ and (e) $Wi = 124800$. The snapshots of the simulation box given on the left illustrate the carbon-carbon bond configuration of the system, where the coloring of the bonds is random.}
    \label{fig:fig2}
\end{figure}
\begin{figure}[H]
    \centering
    \includegraphics[width=0.99\textwidth]{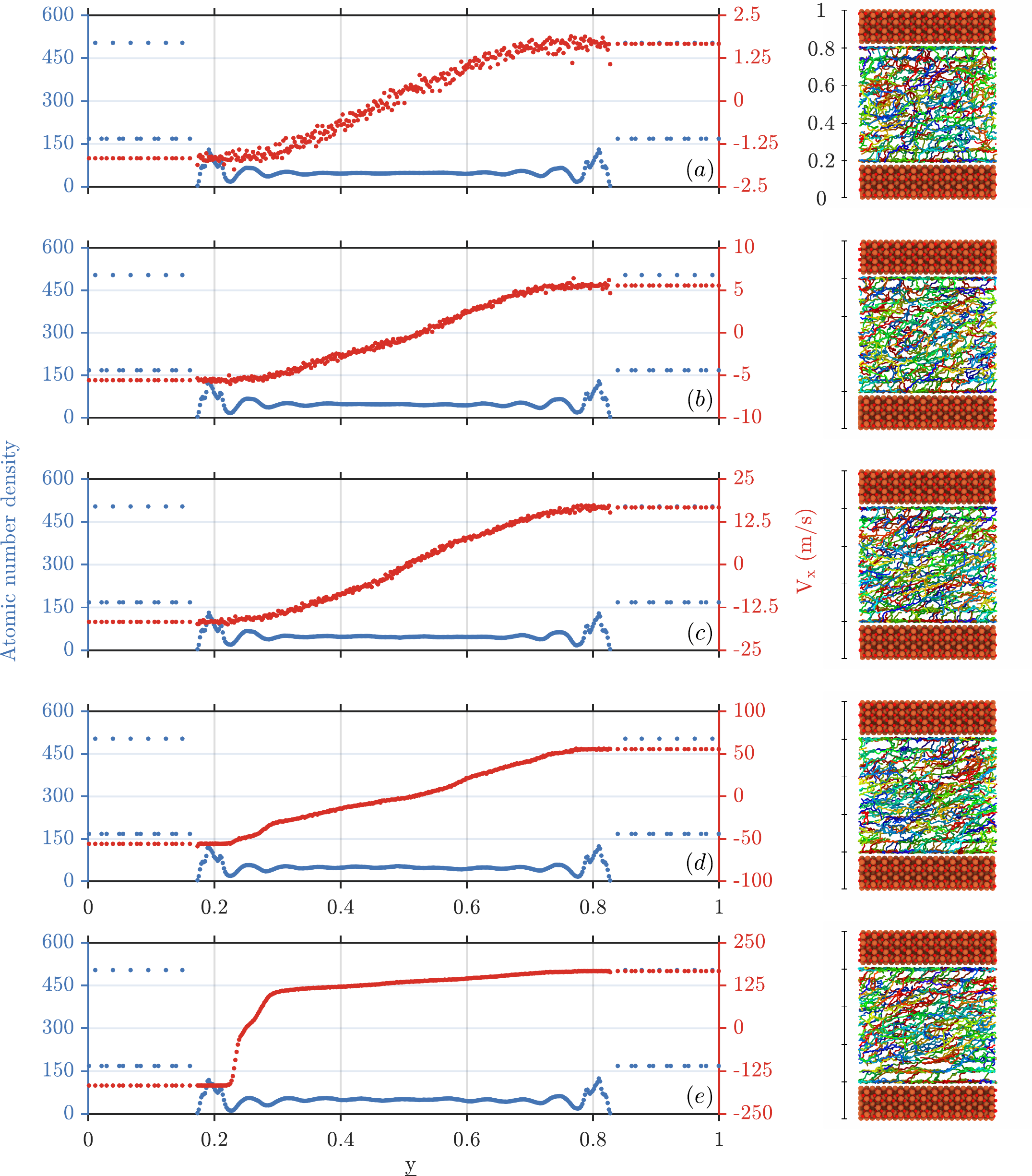}
    \caption{The steady-state spatial configuration and the velocity distribution within the shear of confined star polymer melts where $h = 5.57 \si{\angstrom}$. The shear rate is (a) $\dot{\gamma}=6\times10^8 s^{-1}$, (b) $\dot{\gamma}=6\times10^9 s^{-1}$, (c) $\dot{\gamma}=6\times10^{10} s^{-1}$, (d) $\dot{\gamma}=6\times10^{11} s^{-1}$, and (e) $\dot{\gamma}=6\times10^{12} s^{-1}$ , as a function of the normalized vertical component $\underline{y}$. The corresponding Weissenberg numbers are (a) $Wi = 5.46$, (b) $Wi = 54.6$, (c) $Wi = 546$, (d) $Wi = 5460$, and (e) $Wi = 54600$. The snapshots of the simulation box given on the left illustrate the carbon-carbon bond configuration of the system, where the coloring of the bonds is random.}
    \label{fig:fig3}
\end{figure}

\section{Monomers}
\begin{figure}[H]
    \centering
    \includegraphics[width=0.66\textwidth]{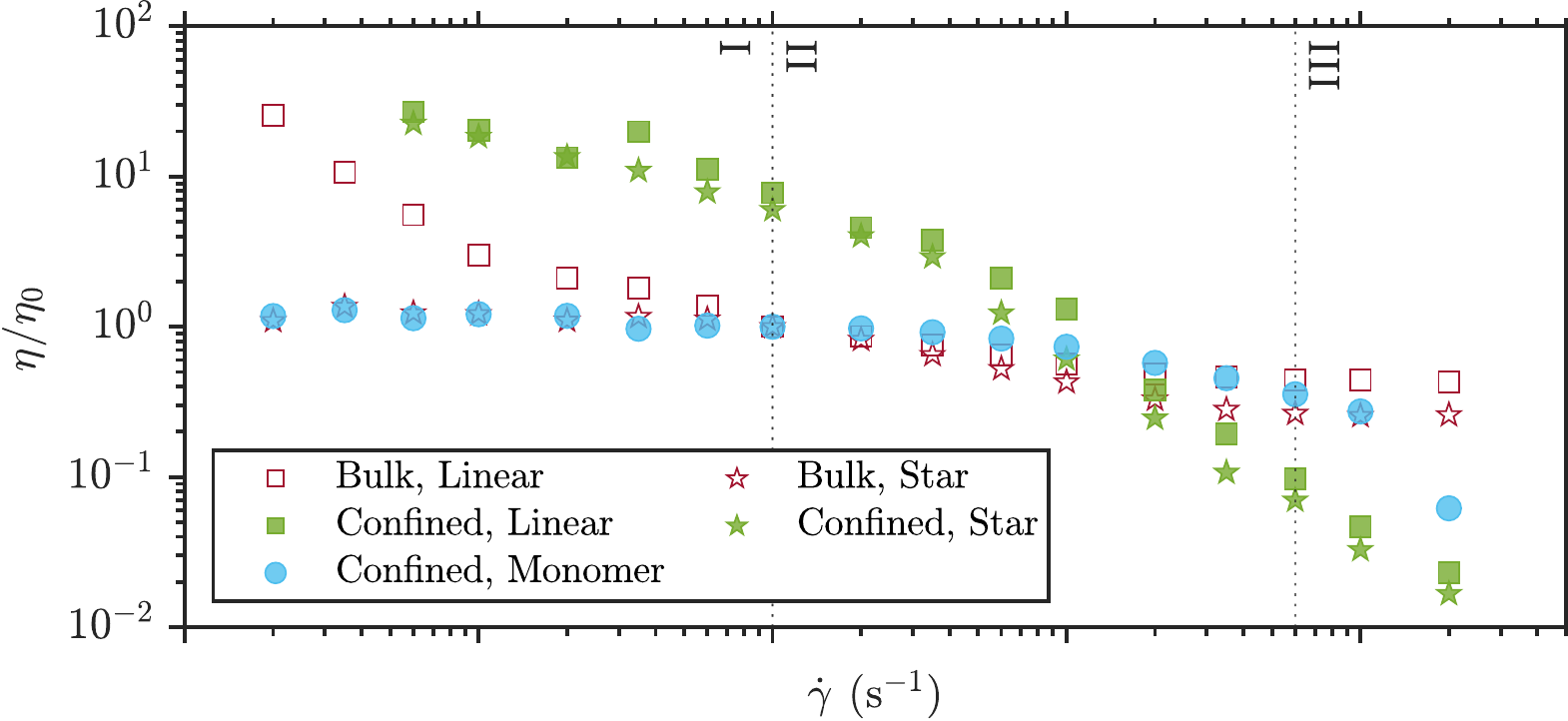}
    \caption{The normalized viscosity of monomers under confinement, obtained by breaking all the carbon-carbon bonds of the linear polymer melt, together with linear and star polymer melts in bulk and under confinement. Linear and star polymers have the unaltered configuration ($k_{\theta} = \epsilon$), whereas $k_{\theta}$ does not apply for the monomers.  \Romannum{1}, \Romannum{2}, and \Romannum{3} indicate the \nth{1} Newtonian plateau, the shear-thinning regime, and the \nth{2} Newtonian plateau for bulk polymers, respectively.} 
    \label{fig:fig4}
\end{figure}

\section{Radius of Gyration}

\begin{figure}[H]
    \centering
    \includegraphics[width=1\textwidth]{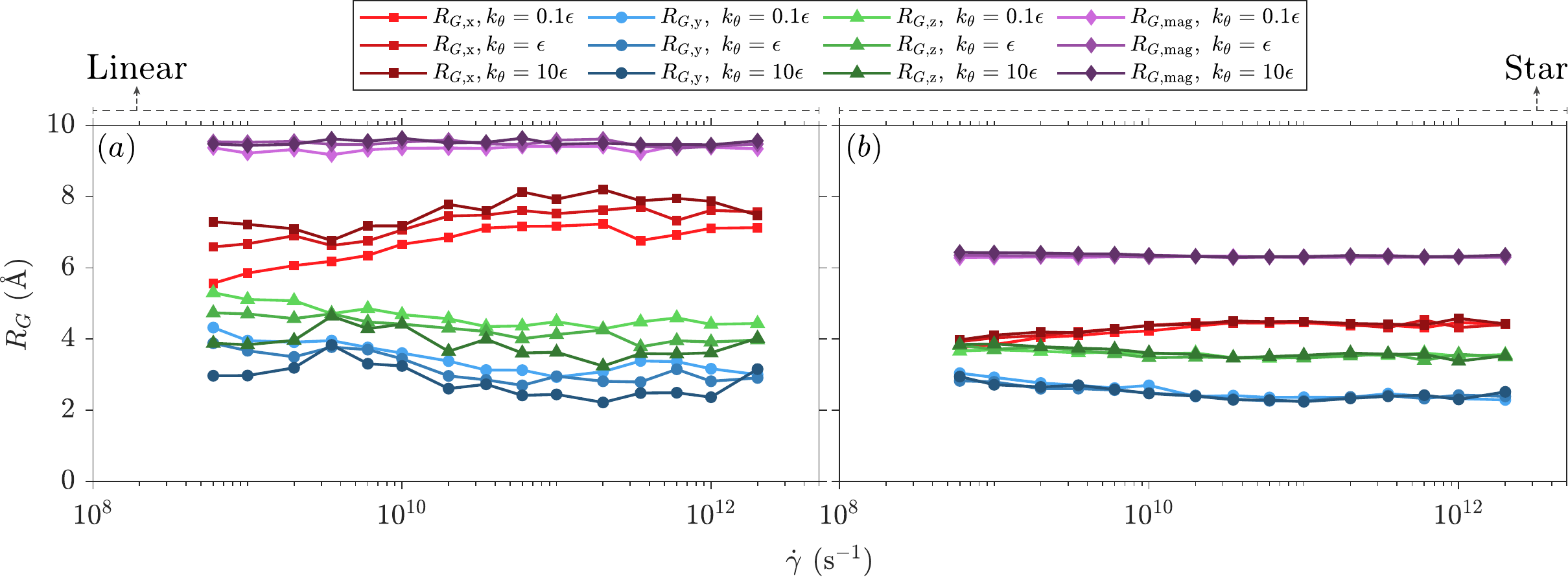}
    \caption{Steady-state radii of gyration (${R_G}$) for confined linear and star polymer melts where $h = 5.57 \si{\angstrom}$. The $x$, $y$, and $z$ components of the radius of gyration, together with the overall magnitude, are given for linear (a) and star (b) polymers. The overall magnitudes $R_{G,\mathrm{mag}}$ do not show significant dependency with $\dot{\gamma}$, the average values read $R_{G,\mathrm{linear}}^{(0.1\epsilon)}=9.34\mathrm{\AA}$,~$R_{G,\mathrm{linear}}^{(\epsilon)}=9.50\mathrm{\AA}$,~$R_{G,\mathrm{linear}}^{(10\epsilon)}=9.52\mathrm{\AA}$;~$R_{G,\mathrm{star}}^{(0.1\epsilon)}=6.32\mathrm{\AA}$,~$R_{G,\mathrm{star}}^{(\epsilon)}=6.31\mathrm{\AA}$, and ~$R_{G,\mathrm{star}}^{(10\epsilon)}=6.36\mathrm{\AA}$.}
    \label{fig:fig5}
\end{figure}

\begin{figure}[H]
    \centering
    \includegraphics[width=0.75\textwidth]{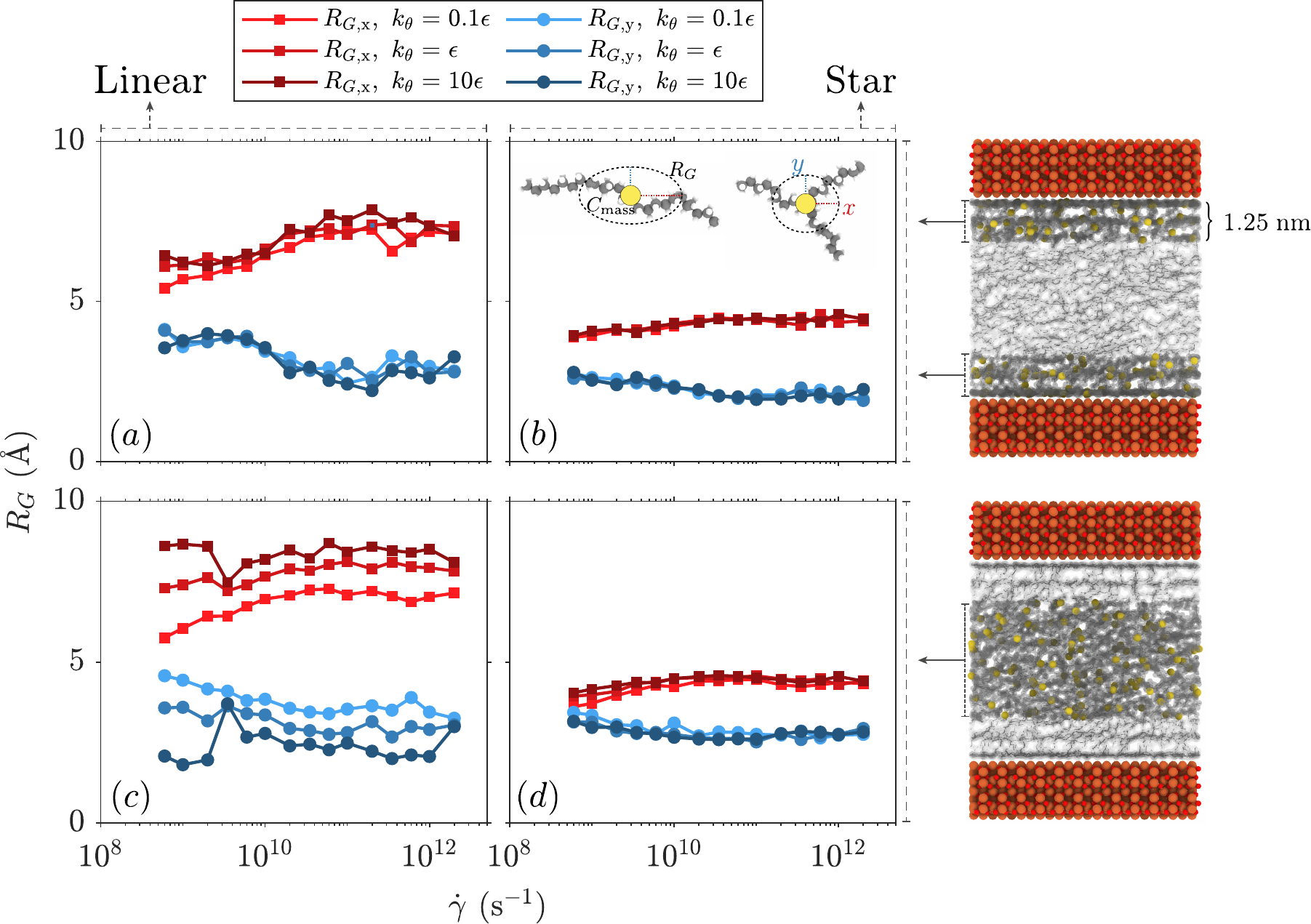}
    \caption{Near/away from the surface steady-state radii of gyration (${R_G}$) for confined linear and star polymer melts where $h = 5.57 \si{\angstrom}$. The $x$ and $y$ components of the radius of gyration are given near the surface for linear (a) and star (b) polymers; away from the surface for linear (c) and star (d) polymers. Especially for star polymers, the relation between viscoelastic load and near-surface dynamics cannot be captured using radii of gyration. The snapshots depict the regions where the components of the radii of gyrations are calculated based on the positions of the center of mass of each chain, whose approximate positions are shown in yellow for a single timestep.}
    \label{fig:fig6}
\end{figure}

\bibliographystyle{h-physrev}

\bibliography{Bib1}